\documentstyle[11pt]{article}
 \newcommand{\be}{\begin{equation}}
\newcommand{\ee}{\end{equation}}
\newcommand{\bea}{\begin{eqnarray}}
\newcommand{\eea}{\end{eqnarray}}
\newcommand{\lb}{\label}

\begin{document}
\begin{flushright}
ZU-TH 24/93
\end{flushright}

\begin{center}
{\bf QUANTUM COSMOLOGY AND THE EMERGENCE OF A CLASSICAL WORLD}
\footnote{ To appear in the Proceedings of the FESt-Workshop on
  Concepts of Space and Time, ed. by E. Rudolph and I.-O. Stamatescu
 (Springer, Berlin, 1993)}
\end{center}
\begin{center}
Claus Kiefer\footnote{Address after October 1, 1993: Institut f\"ur
Theoretische Physik der Universit\"at Freiburg, Hermann-Herder-Str. 3,
 D-79104 Freiburg}\\
Institut f\"ur Theoretische Physik der Universit\"at Z\"urich\\
Sch\"onberggasse 9,
CH-8001 Z\"urich
\end{center}

\vspace{1cm}

\begin{verse}
La pendule s'est arr\^{e}t\'{e}e\\
Personne ne se bouge ... \\
Comme sur les images\\
Il n'y aura plus de nuit

{\em Pierre Reverdy}, from {\em La R\'{e}alit\'{e} Immobile}
\end{verse}

\section{Why quantum cosmology?}

Quantum cosmology is the application of quantum theory to the Universe
 as a whole. At first glance such an attempt seems surprising since
 one is used to apply quantum theory to microscopic systems. Why, then,
 does one wish to extrapolate it to the whole Universe?
This extrapolation is based on the assumption that quantum theory is
{\em universally valid}, in particular that there is no a priori classical
 world. The main motivation for this assumption comes from the kinematical
 non-locality (or non-separability) of quantum theory, i.e. from the
 fact that one cannot in general assign a wave function to a given system
 since it is not isolated but coupled to its natural environment, which
 is again coupled to another environment, and so forth. The extrapolation
 of this quantum entanglement  thus leads inevitably to the concept of a
 wave function for the Universe. Many experiments, notably those which
 contradict Bell's inequalities, have impressively demonstrated this
 fundamental non-separability of quantum theory. As we will see in
 section~3, it is this entanglement between many degrees of freedom which
 is also responsible for the emergence of a classical world. The importance
 of this effect seems to have been overlooked in the traditional discussion
of quantum theory, and led to the belief in an independently existing
 classical world.

Since the dominating interaction in the cosmological realm is gravity,
 this extrapolation immediately has to address the problem of quantizing
 the gravitational field. In fact, since all other known interactions are
 successfully described by quantum field theories it seems unavoidable to
 quantize gravity, too, since the gravitational field is coupled to all
 other fields and it would appear strange, and probably even inconsistent,
 to have a drastically different framework for this one field.  Many
 technical and conceptual difficulties have, however, as yet prevented
 the construction of a consistent and predictive theory of quantum gravity,
 some of which we will briefly describe in the course of this article.

 Quantum cosmology is only meaningful as a physical theory if one can agree
 on how to interpret a universal quantum theory. The traditional
 ``Copenhagen interpretation," for example, strictly denies the idea of
 a fundamental quantum world and assumes from the outset the existence
 of a classical world whose concepts have to be used to interpret the
 results of quantum measurements. Such an attitude appears to be ad hoc
 even in ordinary quantum mechanics since measurement apparata are built
 from atoms which are known to obey the laws of quantum theory. Although
 not being consistent, such a hybrid description of Nature has been
 sanctioned by using purely verbal constructs like ``complementarity."
The conceptual inconsistency of the ``Copenhagen interpretation" becomes
 even more evident in the framework of quantum cosmology, where the whole
Universe including all observers and apparata {\em has to} be described
 in quantum terms from the very beginning. To quote Gell-Mann and Hartle
 (1990): ``Quantum mechanics is best and most fundamentally understood in
 the framework of quantum cosmology."

But since it is impossible to prepare an ensemble of universes, can there
 be any fundamental meaning of a wave function of the Universe? The
 situation can be compared to ordinary quantum field theory where the
 concept of a vacuum state vector is used to derive properties of elementary
 particles but {\em not} to  perform experiments with an ensemble of vacua.
One might expect that in an analogous sense a wave function of the Universe,
 $\Psi$, determines structures of the Universe which can be checked by
 observations. One example is outlined in the last section of this
 article and is concerned with the direction of time. Another example
 is the claim (supported by heuristic considerations only) that $\Psi$
 is peaked around a vanishing cosmological constant (Baum, 1983) leading
 to the {\em prediction} that the cosmological constant {\em is} zero.
 Since such a wave function of the Universe is independent of any observation,
 one must attribute to it the status of reality.

As mentioned above, any attempt to understand quantum cosmology must focus
 on the construction of a quantum theory of gravity. One can distinguish
 several levels of the relationship between quantum theory and gravity
 which we briefly wish to describe. The most fundamental level should be
 described by a quantum theory of all interactions including gravity.
 A promising candidate in recent years has been the theory of superstrings
 which is constructed by using the assumption that the fundamental entities
 are one-dimensional objects instead of local quantum fields. Since many
 mathematical and conceptual problems in this framework have not yet been
 solved, no final agreement on the physical status of this ``theory of
 everything" has been reached so far. A less ambitious level consists in
 the application of formal quantization rules to the general theory of
 relativity. Here one does not attempt to provide a unified theory of all
interactions but tries to focus on the specific quantum aspects of the
 gravitational field. The scale where such aspects are expected to become
 relevant is found by setting the Compton wavelength associated with a mass
 equal to its Schwarzschild radius, and is called the Planck length:
\be l_{Pl} = \sqrt{\frac{\hbar G}{c^3}} \approx 10^{-33}\mbox{cm} .
   \lb{1} \ee
One would expect that the level of quantum general relativity can be
 recovered from superstring theory in an appropriate low-energy limit.
 It might, however, also be the case that quantum general relativity {\em is}
 the fundamental theory, and impressive formal developments have taken place
 recently which support this possibility. We will comment on this below. In
 either case it is justified to study the implications of quantum general
 relativity which is the very conservative framework of tying together
quantum theory and general relativity. Both theories have up to now passed
 all tests in their respective realms.

On the next level we find the framework of quantum field theory for
 non-gravitational fields on a fixed classical background spacetime.
 How this level can be recovered from the more fundamental level of quantum
 general relativity and in particular how the Schr\"odinger equation
 for non-gravitational fields can be derived from quantum gravity, is
 the subject of the third section. On this level one still has no experimental
 test but at least a definite prediction -- the Hawking effect: Black holes
 are not really black if quantum effects are taken into account, but radiate
 with a temperature
\be T=\frac{\hbar c^3}{8\pi Gk_B M} =  \frac{\hbar\kappa}{2\pi ck_B}
     \approx 10^{-6}Kelvin \times \frac{M_{sun}}{M}, \lb{2} \ee
where $M$ is the mass of the black hole, $k_B$ Boltzmann's constant,
 and $\kappa$ the surface gravity of the hole. It is remarkable that
 all fundamental constants of nature appear in this formula. The
significance of the relation (2) for a quantum theory of gravity might
 well turn out to be analogous to the significance of de Broglie's relation
 $p=\hbar k$ for the development of quantum mechanics (Zeh, 1992).

The experimental level of the relation between quantum theory and gravity
 is only reached at the very modest level of the Schr\"odinger equation
 with a Newtonian potential where impressive experiments using neutron
 interferometry have been performed.

What are the main obstacles in constructing a quantum theory of gravity?
 An important formal problem is the non-renormalizability of a perturbative
 expansion of quantum general relativity around a given background spacetime,
 i.e. the fact that an infinite number of parameters would have to be
 determined by experiment to absorb all the infinities of the theory.
 It is therefore not surprising that recent developments mainly focus on
 non-perturbative approaches. A serious obstacle for experiments in this
 field is the extreme smallness of the Planck length (1). One would have
to extrapolate current particle accelerators to the dimensions of the
 Milky Way to be able to probe such a scale -- a hopeless enterprise.
 It is thus hoped that a consistent theory of quantum gravity will reveal
 what the observable effects are. Such effects might well be found in a
 direction not imagined hitherto. One recent investigation, for example, claims
that effects coming from the wave function of the Universe have already been
 observed by the COBE satellite (Salopek, 1993).
 Independent of the lack of experiments, there remain many conceptual problems
 which have to be addressed if gravity has to be quantized. Most of these
 problems are connected in one way or the other with the concept of time.
We will discuss this issue in the next section but will first briefly outline
 the formal framework in which most of the recent investigations are made --
the framework of {\em canonically} quantizing general relativity.

What does it mean to ``quantize" a classical theory? The prominent role in
quantum mechanics is played by the position and momentum operators
 (the ``$p$'s and the $q$'s").
 of a given system. The central property is their non-commutativity which
 gives rise to the famous uncertainty relations. In the Schr\"odinger
 formulation of quantum mechanics one uses wave functions which are
 defined on configuration space, i.e. the space of all position coordinates
 (or, alternatively, on momentum space). ``Canonically quantizing" a classical
 theory now means to identify the ``$p$'s and the $q$'s" of the theory under
 consideration and imposing on them non-trivial commutation relations.
 In the case of the general theory of relativity this involves some preparatory
steps since the $p$'s can only be defined after an appropriate time parameter
 can be distinguished. Since the theory treats all space- and time-coordinates
 on an equal footing one has to   formally rewrite the theory in such a manner
 that time-coordinates appear. This does, of course, not alter the physical
 content of the theory which is invariant under general coordinate
 transformations.

 Basically, the steps in canonically quantizing general relativity are the
 following: The first step consists in foliating spacetime into a family
 of spacelike three-dimensional hypersurfaces -- one decomposes spacetime
into space and time. The metric on these hypersurfaces, $h_{ab}({\bf x})$,
 will play the role of the canonical variable (the $q$).  All quantities
are then decomposed into variables which live on such a
hypersurface and variables which point in the fourth, timelike, dimension.
 It then turns out that the canonical momentum, $\pi^{ab}({\bf x})$, is
 essentially given by the extrinsic curvature of a
three-dimensional
  hypersurface, i.e the quantity which describes
  the embedding of space into spacetime. Loosely speaking, one can say
 that intrinsic and extrinsic geometry are canonically conjugate to each other.

  A major feature of general relativity is its invariance under arbitrary
coordinate transformations. As a consequence one finds that there exist -- at
each space point -- four constraints. Three of them generate coordinate
transformations on the three-dimensional space and are analogous to Gauss' law
in electrodynamics. The fourth,
 so-called {\em
  Hamiltonian  constraint},
 plays a double role: Although being a constraint, it generates the dynamics.
 Its explicit form is
\be {\cal H}\equiv \frac{16\pi G}{c^2} G_{abcd}\pi^{ab}\pi^{cd} -
\frac{c^4}{16\pi G} \sqrt{h}R + {\cal H}_m = 0, \lb{3} \ee
where $\sqrt{h}$ is the square root of the determinant of the three-metric,
 $R$ is the curvature scalar on three-space, and ${\cal H}_m$ is the
Hamiltonian density for non-gravitational fields. The coefficients $G_{abcd}$
depend explicitly on the metric and play itself the role of a metric in the
space of all metrics. The existence of the constraint (3) is directly connected
to the invariance of the theory under reparametrizations of time. It is
therefore not surprising that it is quadratic in the momenta (the same happens,
for example, in the case of the relativistic particle -- invariance under
reparametrizations of the world line parameter leads to the constraint $p^2
+m^2=0$). The presence of the invariance under coordinate transformations
 and its associated constraints lies at the heart of the quantization problem.
To quote Pauli (1955):
\begin{quote} Es scheint mir \ldots, da\ss\ nicht so sehr die Linearit\"at oder
Nichtlinearit\"at Kern der Sache ist, sondern eben der Umstand, da\ss\ hier
eine allgemeinere Gruppe als die Lorentzgruppe vorhanden ist \ldots
\footnote{It seems to me \ldots that it is not so much the linearity or
non-linearity which forms the heart of the matter, but the very fact that here
a more general group than the Lorentz group is present \ldots }
\end {quote}

Quantization now proceeds, at least formally,  by elevating the metric and its
 momentum to the status of operators and imposing the commutation relations
  \be [h_{ab}({\bf x}), \pi^{cd}({\bf y})] =i\hbar \delta^c_{(a}\delta^d_{b)}
  \delta({\bf x},{\bf y}) \lb{4} \ee
 in full analogy to the commutation relation $[q,p]=i\hbar$ in quantum
mechanics.
 One specific realization of (4) is provided by the substitution (in analogy to
substituting $p\to \frac{\hbar}{i}\frac{d}{dq}$)
\be  \pi^{ab} \to \frac{\hbar}{i} \frac{\delta}{\delta h_{ab}}. \lb{5} \ee
The classical constraint (3) is then formally implemented in the quantum
theory by inserting (4) and (5) into (3) and applying it on wave functionals
 $\Psi$ which depend -- apart from non-gravitational fields denoted by
 $\phi$ --  on the three-metric, i.e.
\be {\cal H}\Psi[h_{ab}({\bf x}),\phi({\bf x})] =
   \left(- \frac{16\pi\hbar^2 G}{c^2} G_{abcd} \frac{\delta^2}{\delta h_{ab}
 \delta h_{cd}}
   -\frac{c^4}{16\pi G}\sqrt{h}R +{\cal H}_m
\right)\Psi  = 0. \lb{6} \ee
This equation is called the {\em Wheeler-DeWitt equation} in honour of the
 pioneering work of DeWitt (1967) and Wheeler (1968). It has the form of a
 zero-energy Schr\"odinger equation. There are of course many technical
 problems such as factor ordering or regularization which we will not be
 able to address in this article (see, e.g., Kucha\v{r}, 1992).

The quantization of the remaining three constraints leads to the condition that
this wave functional actually does not change under a coordinate transformation
of the three-metric but is a function of the {\em geometry} only. The
configuration space is thus the space of all three-geometries and is called
{\em superspace}. An important physical consequence of the commutation rules
(4) is the ``uncertainty" between the three-dimensional space and its embedding
in the fourth dimension: The concept of spacetime is a classical concept only
with no fundamental meaning in quantum theory. This is fully analogous to the
concept of a particle trajectory which has no fundamental meaning in quantum
mechanics due to the uncertainty between position and momentum. When applied to
cosmology, the wave functional $\Psi$ in (6) is the desired ``wave function of
the Universe" which has been mentioned at the beginning of this section.

An important development has been the discovery of appropriate variables which
enables one to find exact formal solutions of Eq. (6) in the absence of matter
(Ashtekar, 1991). This became possible since the complicated potential term of
(6) disappears when the equation is written in terms of the new variables,
which have a strong similarity to Yang-Mills variables. The solutions can be
classified in terms of loops and knots and exhibit an interesting structure of
space, of which may be the most important is the existence of a minimal length
of the order of the Planck length (1).  Consequently, smaller scales do not
have any operational meaning.

We have not yet addressed the issue of boundary conditions to be imposed on the
Wheeler-DeWitt equation. In contrast to systems in the laboratory, they are not
at our disposal. In fact, the question of boundary conditions has been of great
interest in recent years, basically because of the {\em no boundary proposal}
by Hartle and Hawking (1983). These two authors express the wave functional as
a formal path integral where the sum is over euclidean (instead of lorentzian)
geometries. The no boundary proposal then consists in the condition that one
performs a sum over compact manifolds with one boundary only -- the boundary
which is given by the considered universe. The lack of a second boundary (for
example at a small size of the universe) saves one from the need to find
appropriate boundary conditions there.  Unfortunately, these path integrals can
only be evaluated in a semiclassical approximation and, moreover, do not lead
to a unique wave function. We will return to the question of boundary
conditions in

the last section in connection with a discussion of the arrow of time.

Since the full equation (6) is in general difficult to handle, most
applications have been performed in a restricted framework where only finitely
many degrees of freedom are quantized. A typical example in the cosmological
context is the quantization of the scale factor $a$ of a Friedmann Universe. If
in addition a (conformally coupled) scalar field $\phi$ is taken into account
to simulate the matter content of the Universe we find instead of (6) the much
simpler equation
\be H\psi(a,\phi) = \left(\frac{\partial^2}{\partial a^2} -
 \frac{\partial^2}{\partial\phi^2} - a^2 +\phi^2\right)\psi(a,\phi) =0.
\lb{7} \ee
This equation has the form of an indefinite harmonic oscillator and can serve
as a useful tool in studying conceptual questions in quantum cosmology (see the
following sections).

The canonical quantization scheme outlined in this section still contains the
structure of a differentiable three-dimensional manifold. There exist more
ambitious approaches to quantum gravity which give up even this structure and
start with a quantization of topology, see e.g. Isham, Kubyshin and Renteln
(1990) who construct wave functions which live on a family of topologies. We
will not, however, follow these approaches any further.

\section{Time in quantum gravity}
In the preface to Max Jammer's book on the {\em Problem of Space}, Einstein
writes
\begin{quote}
Es hat schweren Ringens bedurft, um zu dem f\"ur die theoretische Entwicklung
unentbehrlichen Begriff des selbst\"andigen und absoluten Raumes zu gelangen.
Und es hat nicht geringerer Anstrengung bedurft, um diesen Begriff
nachtr\"aglich wieder zu \"uberwinden -- ein Proze\ss , der wahrscheinlich noch
keineswegs beendet ist.\footnote{It was a hard struggle to gain the concept of
independent and absolute space which is indispensible for the theoretical
development. And it has not been a smaller effort to overcome this concept
later on, a process which probably has not yet come to an end.}
\end{quote}
Although Einstein refers in this quotation to space only, the same can be said
about time. Absolute time, as well as absolute space, was an indispensable
ingredient in Newton's theory of motion and more powerful in the development of
dynamics than, for example, the notion of relative time which was put forward
by Leibniz. Time kept its absolute status even in non-relativistic quantum
mechanics -- the $t$ in Schr\"odinger's equation is still an external parameter
and not turned into a quantum operator.

Although the causal structure of spacetime has drastically changed with the
advent of the special theory of relativity by dropping the notion of absolute
simultaneity, spacetime is there still understood as a {\em non}-dynamical
entity which provides an arena for the laws of physics but does not take part
in the play. There is hence still no quantum operator for time in relativistic
quantum field theory, although one now has the possibility to choose any
spacelike hypersurface as a time parameter.
These spacelike hypersurfaces can be deformed independently at each space
point, which gives rise to a local or
 ``many-fingered" time $\tau({\bf x})$ instead of one single $t$. This local
time appears also in the field theoretic
 Schr\"odinger equation, which is an equation for wave {\em functionals}
depending on fields $\phi({\bf x})$ (which play the role of the ``$q$'s" in
quantum mechanics). This equation reads
\be i\hbar\frac{\delta\psi[\phi({\bf x})]} {\delta\tau({\bf x})} =
     {\cal H}_m\psi[\phi({\bf x})], \lb{8} \ee
where
 ${\cal H}_m$ is the Hamiltonian density connected with  $\phi({\bf x})$.

Spacetime becomes a {\em dynamical} object, in analogy to a particle
trajectory, only in the general theory of relativity where the spacetime metric
describes the gravitational field, which is subject to Einstein's field
equations. The quantization of gravity therefore unavoidably has to address the
quantization of time. Why would one expect any problems to occur in this step?
One has to recall that the presence of an external time is an essential
ingredient of quantum mechanics -- matrix elements are calculated at a given
time, and measurements are performed at a given time. Moreover, probabilities
are preserved {\em in} time.  The {\em absence} of any time parameter is a
fundamental property of the Wheeler-DeWitt equation (6). This is not surprising
since, as we have argued above, the concept of spacetime has no fundamental
meaning in quantum gravity.  The question therefore arises: Can one introduce a
viable concept of time on the fundamental level of quantum gravity itself or
only in a semiclassical appr

oximation? A critical investigation into the concept of time may lead to
fruitful insights into the structure of the desired theory. To quote Einstein
again:
\begin{quote}
\ldots und doch ist es im Interesse der Wissenschaft n\"otig, da\ss\ immer
wieder an diesen fundamentalen Begriffen Kritik ge\"ubt wird, damit wir nicht
unwissentlich von ihnen beherrscht werden.\footnote{\ldots and yet it is
necessary, in the interest of science, to call these fundamental concepts again
and again into question so that we are not governed by them without realising
it.}
\end{quote}
The emergence of a semiclassical time from quantum gravity will be discussed in
the next section. Here we focus on the level of Equation (6) itself. An
important property of the Wheeler-DeWitt equation is its hyperbolic nature,
i.e. its behaviour as a wave equation. This can be recognized from the
coefficients $G_{abcd}$ by treating them -- at each space point separately --
as the
elements of a symmetric $6\times 6$ matrix which after diagonalization has the
signature $(-,+,+,+,+,+)$. It is important to note that there remains {\em one}
global minus sign after gauge degrees of freedom have been eliminated.
This minus sign is basically given by the size of the Universe, which may thus
be considered as an {\em intrinsic time} variable. $\Psi$ has now to be
interpreted as a probability amplitude {\em for} time, but not {\em in} time.
All other degrees of freedom, including physical clocks, are correlated with
intrinsic time. The hyperbolic nature of (6) allows the formulation of a Cauchy
problem with respect to this time.  In spite of its static appearance, the
Wheeler-DeWitt equation describes an intrinsic dynamics! This has drastic
consequences for the behaviour of wave packets in the case of a recollapsing
universe (Zeh, 1988). In the classical theory, the recollapsing leg of the
history of the universe
 can be considered as the deterministic successor of the expanding leg. This is
no longer true in quantum cosmology! The wave equation (6) describes dynamics
with respect to intrinsic time, which in simple models like (7) is the radius
of the universe. The expanding and recollapsing legs of a wave packet
concentrated near the classical trajectory in configuration space can thus not
be distinguished intrinsically. With respect to the Cauchy problem the
returning wave packet must be present initially.

This can already be seen in the simple model (7). To construct a wave tube
which follows the collapsing trajectory, one must use solutions to (7) which
fall off for large values of $a$ and $\phi$. This forces one to use the
normalisable harmonic oscillator eigenfunctions in a wave packet solution which
thus reads
\be \psi(a,\phi) = \sum_n A_n \frac{H_n(a)H_n(\phi)}{2^n n!} \exp\left(
     -\frac{1}{2}a^2 - \frac{1}{2}\phi^2\right), \lb{9} \ee
where $A_n$ are coefficients which are peaked around some $n=\bar{n}$, and
$H_n$ are the Hermite polynomials. The solution (9) automatically describes
 {\em two} packets at $a=0$ (see figure~1). While in the simple oscillator
model (7) there is no dispersion of the wave packet, this is no longer true in
more realistic examples like the case of a massive scalar field in a Friedmann
universe (Kiefer, 1988; Zeh, 1992).  The demand for the wave packet to go to
zero at large radii (otherwise it cannot correspond to a recollapsing universe)
unavoidably leads to a {\em smearing} of the packet in regions close to the
classical turning point. This demonstrates that a WKB approximation cannot hold
 in the whole region of an expanding and recollapsing trajectory in
configuration space. This has important consequences for the discussion of the
arrow of time (see the last section).

The above considerations are at present only of a heuristic nature. It is still
unclear what the Hilbert space structure of quantum theory is, if it is
necessary at all. Imposing the Schr\"odinger inner product onto the solutions
of (6) in general leads to a diverging result if {\em all} variables are
integrated over. Recalling the hyperbolic structure of this equation, it would
seem to be more appropriate to choose a Klein-Gordon inner product like in
relativistic quantum mechanics and to integrate over ``spacelike" hypersurfaces
$a=constant$.  Unlike relativistic quantum mechanics it is, however, here not
possible to make a decomposition into positive and negative frequencies (see
Kucha\v{r}, 1992). This has prompted some authors to invoke a third
quantization of the theory, i.e. to elevate the wave functional $\Psi$ to an
operator in some ``new" Hilbert space. We will, however, not follow these
approaches any further. We also mention that there are attempts which try to
{\em first} solve the classical con

straint (3) and only {\em then} make the transition to quantum theory. This
creates its own problems, and we refer to the excellent review articles by
Isham (1992) and Kucha\v{r} (1992) for  details.

\section{Decoherence and the recovery of the Schr\"o\-ding\-er equation}

We now address the issue of how the level of quantum field theory in a
classical spacetime background can be recovered from quantum gravity where
there is no spacetime. This will basically involve two steps. The first step is
the derivation of the Schr\"odinger equation (8) from the Wheeler-DeWitt
equation (6). The second step is to understand the unobservability of
non-classical states for the gravitational field. A detailed review of these
steps can be found, e.g., in Kiefer (1993b).

The basic observation which goes into the development of the first step is the
fact that the length scale contained in (6), the Planck length (1), is much
smaller than any relevant scale of non-gravitational physics. This enables one
to make a formal expansion of the wave functional in (6) in powers of the
Planck length (or, equivalently, the gravitational constant).
If there were no non-gravitational fields in (6), an expansion with respect to
$G$ would be fully equivalent to an expansion in powers of $\hbar$ which is the
usual semiclassical (``WKB") expansion, since both constants would appear in
the combination $G\hbar$ only. As far as gravity is concerned, the present
expansion scheme thus {\em is} a WKB expansion. This is no longer true for
other fields in which case the situation is analogous to a Born-Oppenheimer
approximation in molecular physics: The gravitational part in (6) corresponds
to the heavy nuclei whose kinetic terms are neglected in a first approximation
while the remaining part corresponds to the light electrons. To highest order,
the wave functional   depends only on the gravitational field,
\be \Psi_0 = C[h_{ab}]\exp\left(\frac{ic^2}{32\pi G\hbar}S_0[h_{ab}]
\right), \lb{10} \ee
where $S_0$ obeys the Hamilton-Jacobi equation for gravity. This equation is
equivalent to all of Einstein's field equations and describes a classical
gravitational background in the sense that one can assign classical
``trajectories" to it.  Each trajectory, which represents a whole spacetime,
runs orthogonally to hyperspaces $S_0=constant$ in configuration space.
Formally, this is the same as the recovery of geometrical optics from Maxwell's
equations.

In the next order of approximation the wave functional assumes the form
\be \Psi_1= \Psi_0\ \chi[h_{ab},\phi], \lb{11} \ee
where the wave functional $\chi$ also depends on non-gravitational fields and
obeys the equation (Banks, 1985)
\be i\hbar G_{abcd}\frac{\delta S_0}{\delta h_{ab}} \frac{\delta\chi}
  {\delta h_{cd}} \equiv i\hbar \frac{\delta\chi}{\delta\tau({\bf x})}
  ={\cal H}_m\chi. \lb{12} \ee
This is nothing else but the functional Schr\"odinger equation (8) for
non-gravita\-tio\-nal fields propagating on {\em one} of the classical
spacetimes described by $S_0$. This spacetime is parametrized by the
``many-fingered time"
 $\tau({\bf x})$ appearing in (12).

As Julian Barbour (1992) emphasized, this WKB-time corresponds exactly to the
notion of {\em ephemeris time} used by astronomers. Ephemeris time is the
extraction of time, in retrospect, from actual observations of celestial
bodies, see for example Clemence (1957).  The semiclassical time in (12) is
thus defined by the actual motion of bodies in the real world. It is amazing
that this exactly corresponds to the concept of time used by the ancient Greeks
who defined time by the motion of the celestial bodies (already Plato used the
term ephemeris time). It is clear that there can be no emergence of a
semiclassical time for flat Minkowski space which demonstrates the absence of
any concept of absolute time.  It is the three-dimensional geometry which
carries information about time, see Baierlein, Sharp and Wheeler (1962). To
determine ephemeris time, eventually all available information about motion in
the Universe has to be taken into account. An impressive example is the case of
the binary pulsar PSR 1913+16

. Using general relativity, ephemeris time can only be consistently extracted
from the orbital motion, if the gravitational pull of the whole Galaxy on the
pulsar is taken into account, as was demonstrated by Damour and Taylor (1991).

If one proceeds with the above approximation scheme to the next order (Kiefer
and Singh, 1991), one can derive correction terms to the Schr\"odinger equation
(12) which are proportional to the gravitational constant. In addition one
finds a back reaction of the non-gravitational fields onto the gravitational
background which modifies the definition of semiclassical time. One can
calculate concrete results from these correction terms, such as the quantum
gravitational correction to the trace anomaly in De~Sitter space (Kiefer,
1993b).

Is the recovery of the Schr\"odinger equation in a classical spacetime
sufficient for the understanding of the classical behaviour of the spacetime
geometry in our world? The answer must be {\em no} since one still has to focus
on the issue of superpositions of different ``classical" states of the
gravitational field. This problem is even of direct relevance in the above
derivation: If one takes a superposition of two semiclassical states, for
example the state (10) and its complex conjugate, one {\em cannot} recover the
Schr\"odinger equation (Barbour, 1993). This equation follows only if a
special, {\em complex}, state like (10) is taken as the starting point. As one
recognizes immediately, this issue is directly connected with the emergence of
the $i$ and the use of complex wave functions in ordinary quantum theory -- the
$i$ in the Schr\"odinger equation is taken directly from the state (10) of the
gravitational field. How, then, can one justify the use of such a special
state? A possible answer is provi

ded by the mechanism of decoherence (Kiefer, 1993a). The key ingredient is the
quantum entanglement of the wave function of the Universe, i.e. the existence
of correlations between a large number of degrees of freedom, which we
discussed in the first section. Since only very few degrees of freedom in this
wave function are accessible to observations, the relevant object is {\em not}
the full wave function but the reduced density matrix which is obtained by
tracing out all irrelevant (unobservable) degrees of freedom. If the only
relevant degree of freedom were the scale factor, this density matrix would
read
\be \rho(a,a') = \mbox{Tr}_{\phi}\Psi^*[a',\phi]\Psi[a,\phi], \lb{13} \ee
where $\phi$ stands for the irrelevant degrees of freedom. These can be
gravitational degrees of freedom like gravitational waves as well as, for
example, matter density perturbations. They ``measure" the gravitational
background and force it to become classical (Zeh, 1986; Kiefer, 1987). In this
way ``quasiclassical domains" emerge from the fundamental quantum world
(Gell-Mann and Hartle, 1990). When applied to simple models, the tracing out of
such irrelevant degrees of freedom suppresses interference terms between
different WKB components of the form (10). A conformally coupled scalar field
in a Friedmann universe, for example, leads to a suppression factor of the
interference between (10) and its complex conjugate given by
\be \exp\left(-\frac{\pi m H_0^2 a^3}{128}\right), \lb{14} \ee
where $m$, $H_0$, and $a$ are, respectively, the mass of the scalar field, the
Hubble parameter, and the scale factor (Kiefer, 1992).
 This is very tiny, except for small radii of the Universe and near the region
of the classical turning point where the Hubble parameter vanishes. Quantum
gravity itself thus contains the seeds for the emergence of a classical
geometry, but also describes its limit.

\section{The direction of time}

It is an obvious fact that most phenomena in Nature distinguish a direction of
time (see, for example, Zeh, 1992): Electromagnetic waves are observed in their
retarded form only, where the fields causally follow from their sources. The
increase of entropy, as it is expressed in the second law of thermodynamics,
also defines a time direction. This is directly connected with the
psychological arrow of time -- we remember the past but not the future. In
quantum mechanics it is the irreversible measurement process and in cosmology
the expansion of the Universe, as well as the local growing of inhomogeneities,
which determine a direction of time.

And yet, the fundamental laws of physics are invariant under time reversal (the
only exception being the small CP violation in weak interactions). How, then,
can one understand that most phenomena distinguish a direction of time? The
answer lies in the possibility of {\em very special} boundary conditions such
as an initial condition of low entropy.

 One assumes now that the occurrence of such boundary conditions  can be
understood within the dynamical laws of physics and that it makes sense to
search for a common root of the various arrows of time -- the {\em master arrow
of time}.  Such an assumption transcends the Newtonian separation into laws and
boundary conditions by also seeking physical explanations for the latter.

Where lies the key to the understanding of the irreversibility of time? As in
particular Penrose (1979) has convincingly emphasized, it is primarily the high
unoccupied entropy capacity of the gravitational field which allows for the
emergence of structure far from thermodynamical equilibrium. Whereas the
non-gravitational part of the entropy reaches its maximum for a homogeneous
state, the opposite is true for the gravitational part which tries to develop a
highly clumpy state. It is therefore a cosmological problem to justify the
presence of an initial state of very low gravitational entropy, i.e. a very
homogeneous state. This has provoked Penrose to formulate his {\em Weyl tensor
hypothesis} that the Weyl tensor vanishes at singularities in the past but not
at those in the future. The Weyl tensor is that part of the Riemann tensor
which is not fixed by the field equations (in which only the Ricci tensor
enters) but by the boundary conditions only. It describes the degrees of
freedom of the gravitational

field. Since it vanishes exactly for a homogeneous and isotropic Friedmann
Universe, it can be taken as a heuristic measure for inhomogeneity and,
therefore, for gravitational entropy.

The arrow of time can of course only be explained by the Weyl tensor
hypothe\-sis if it can be derived from some fundamental theory. As we have
argued in the previous sections, the fundamental framework to address such
questions is quantum cosmology. In the configuration space for the wave
function of the Universe there is no intrinsic distinction between Big Bang and
Big Crunch since both just correspond to regions of low scale factor $a$. The
desired boundary condition is then a boundary condition for the wave function
at small scales. It has therefore been suggested (Zeh, 1993) to impose the
boundary condition that for small scales the wave function depends only on $a$
 but {\em not} on any other degrees of freedom. These degrees of freedom emerge
only with increasing $a$ when they are in their ground state, as can be seen
from a discussion of the Wheeler-DeWitt equation (Conradi, 1992). A similar
behaviour was also derived from the no boundary condition (Hawking, 1985). At
least heuristically, this impl

ements a low gravitational entropy at small scales. Complexity, and therefore
entropy, increases with increasing scale factor. Since more and more degrees of
freedom come into play, decoherence also increases and the Universe becomes
more and more classical.  The thermodynamical arrow of time is thus
inextricably tied to the cosmological arrow of time.

Such a boundary condition has important physical consequences. Since the
thermodynamical arrow of time is correlated with the scale factor but not with
any classical trajectory (which is absent in quantum cosmology), this would
mean that in the case of a closed universe time would formally reverse its
direction near the turning point. We call such a reverse formal since no
information gathering system could survive this reversal. All observers in any
branch of the wave function would consider their universe as expanding. This
boundary condition would also have drastic consequences for the behaviour of
black holes (Zeh, 1993) since they would formally become white holes during the
recontraction phase, and the formation of a horizon would be prevented. The
whole quantum universe would be singularity free in this case.

These considerations are of course still speculative. They are, however, only
based on the established formal framework of quantum theory and cosmology. They
demonstrate that quantum cosmology, if taken seriously, can yield a picture of
the Universe which drastically modifies the classical Big Bang model.

\begin{center}
{\bf Acknowledgement}
\end{center}
I am grateful to H.-Dieter Zeh for many stimulating discussions and a
critical reading of this manuscript. I also thank Claudia Pertzborn and
 Tejinder Singh for their comments on this manuscript.

\vspace{2cm}

\begin{center}
{\bf References}
\end{center}

\begin{description}
\item[A. Ashtekar] (1991): {\em Lectures on non-perturbative canonical gravity}
(World Scien\-tific, Singapore).
\item[R. G. Baierlein, D. H. Sharp, and J. A. Wheeler] (1962):
Three-Dimen\-sio\-nal Geometry as Carrier of Information about Time. Phys. Rev.
{\bf 126}, 1864.
\item[T. Banks] (1985): TCP, Quantum Gravity, the Cosmological Constant, and
all that \ldots . Nucl. Phys. {\bf B249}, 332.
\item[J. B. Barbour] (1992): Personal communication.
\item[J. B. Barbour] (1993): Time and complex numbers in canonical quantum
gravity. Phys. Rev. D {\bf 47}, 5422.
\item[E. Baum] (1983): Zero cosmological constant from minimum action. Phys.
Lett. {\bf B133}, 185.
\item[G. M. Clemence] (1957): Astronomical Time. Rev. Mod. Phys. {\bf 29}, 2.
\item[H. D. Conradi] (1992): Initial state in quantum cosmology. Phys. Rev. D
{\bf 46}, 612.
\item[T. Damour and J. H. Taylor] (1991): On the orbital period change of the
binary pulsar PSR 1913+16. Astrophys. Journal {\bf 366}, 501.
\item[B. S. DeWitt] (1967): Quantum Theory of Gravity I. The Canonical Theory.
Phys. Rev. {\bf 160}, 1113.
\item[M. Gell-Mann and J. B. Hartle] (1990): ``Quantum Mechanics in the Light
of Quantum Cosmology", In {\em Complexity, Entropy and the Physics of
Information}, ed. by W. H. Zurek (Addison Wesley).
\item[J. B. Hartle and S. W. Hawking] (1983): Wave Function of the Universe.
Phys. Rev. D {\bf 28}, 2960.
\item[S. W. Hawking] (1985): Arrow of Time in Cosmology. Phys. Rev. D {\bf 32},
2489.
\item[C. J. Isham] (1992): ``Canonical Quantum Gravity and the Problem of
Time", Lectures presented at the NATO Advanced Study Institute {\em Recent
problems in Mathematical Physics}, Salamanca, June 15-27, 1992.
\item[C. J. Isham, Y. Kubyshin, and R. Renteln] (1990): Quantum norm theory and
the quantisation of metric topology. Class. Quantum Grav. {\bf 7}, 1053.
\item[C. Kiefer] (1987): Continuous measurement of mini-superspace variables by
higher multipoles. Class. Quantum Grav. {\bf 4}, 1369.
\item[C. Kiefer] (1988): Wave packets in minisuperspace. Phys. Rev. D {\bf 38},
1761.
\item[C. Kiefer] (1992): Decoherence in quantum electrodynamics and quantum
gravity. Phys. Rev. D {\bf 46}, 1658.
\item[C. Kiefer] (1993a): Topology, decoherence, and semiclassical gravity.
Phys. Rev. D {\bf 47}, 5414.
\item[C. Kiefer] (1993b): ``The Semiclassical Approximation to Quantum
Gravity", to appear in {\em The Canonical Formalism in Classical and Quantum
General Relativity} (Springer, Berlin).
\item[C. Kiefer and T. P. Singh] (1991): Quantum Gravitational Corrections to
the Functional Schr\"odinger Equation. Phys. Rev. D {\bf 44}, 1067.
\item[K. V. Kucha\v{r}] (1992): ``Time and Interpretations of Quantum Gravity",
In {\em Proceedings of the 4th Canadian Conference on General Relativity and
Relativistic Astrophysics}, ed. by G. Kunstatter, D. Vincent and J. Williams
(World Scientific, Singapore).
\item[W. Pauli] (1955): Schlu\ss wort. Helv. Phys. Acta Suppl. {\bf 4}, 266.
\item[R. Penrose] (1979): ``Singularities and Time-Asymmetry", In {\em General
Relativity}, ed. by S. W. Hawking and W. Israel (Cambridge University Press,
Cambridge).
\item[D. S. Salopek] (1993): ``Searching for Quantum Gravity Effects in
Cosmological Data", to appear in the {\em Proceedings of Les Journ\'{e}es
Relativistes}, Brussels, April 5-7, 1993 (World Scientific, Singapore).
\item[J. A. Wheeler] (1968): ``Superspace and the nature of quantum
geometrodynamics", In {\em Battelle rencontres}, ed. by C. M. DeWitt and J. A.
Wheeler (Benjamin).
\item[H. D. Zeh] (1986): Emergence of Classical Time from a Universal Wave
Function. Phys. Lett. {\bf A116}, 9.
\item[H. D. Zeh] (1988): Time in Quantum Gravity. Phys. Lett. {\bf A126}, 311.
\item[H. D. Zeh] (1992): {\em The Physical Basis of The Direction of Time}
(Springer, Berlin).
\item[H. D. Zeh] (1993): ``Time (a-)symmetry in a recollapsing universe", In
{\em Physical Origin of Time Asymmetry}, ed. by J. J. Halliwell, J.
Perez-Mercader and W. H. Zurek (Cambridge University Press, Cambridge).

\end{description}

\vspace{2cm}

\begin{center}
{\bf Figure Caption}
\end{center}
Wave packet solution of the Wheeler-DeWitt equation (7).

\end{document}